\documentclass[preprint, prX]{revtex4-1} 

\usepackage{hyperref}
\usepackage{amssymb}
\usepackage [dvips]{graphicx}
 \usepackage{graphicx}\usepackage{color}
 \usepackage {amsmath}
\usepackage{array}

\usepackage{dcolumn}%
\usepackage{bm}%

\begin{document}  

\title{Novel method to obtain the anisotropic parameter of microcells}
\author{Emiliano Ter\'an-Bobadilla}
\email{eteran@uas.edu.mx}
\affiliation{Facultad de Ciencias F\'isico-Matem\'aticas\\ Universidad Aut\'onoma de Sinaloa}

\author{Eugenio Rafael M\'endez M\'endez}
\affiliation{Departamento de \'Optica\\ Centro de Investigaci\'on Cient\'ifica y de Educaci\'on Superior de Ensenada}
\date{\today}

\begin{abstract}

In this paper we present a technique to measure the anisotropy parameter a suspension of microscopic particles. To estimate this parameter we use an integrating sphere as light collector, as well as a powder sample spherical particles of TiO$ _2 $ with an average diameter $ a = 0.25 \ \mu $m. The results for the sample of TiO$ _2$ are consistent with Mie calculations.
\end{abstract}
\maketitle


\section{Introduction\label{sec:intro}}

The anisotropy parameter of a system of particles is defined as the average scattering angle, or as the first moment of the phase function of the system. The phase function of a system represents the probability density function of the scattered light. Therefore, the measurement of this parameter involves the implementation of an experimental arrangement that captures all the light scattered or as we will see, a ratio of the reflected and transmitted light. There are different methodologies to estimate this parameter, perhaps the best known is the method of scatterometer.

The scatterometer is an optical instrument that allows us to estimate the scattered light as a function of scattering angle. However, one of its main drawbacks is that you need to implement a substantial infrastructure, plus a careful interpretation of the results. On the other hand, the method of integrating spheres allows us to implement a highly versatile experimental setup to measure this parameter.

The integrating sphere method is commonly used to determine the optical properties of biological systems, its implementation usually requires the use of two integrating sphere. In the proposed settlement of this work will be done only using an integrating sphere and will be focused on the characterization of microparticles.

The estimation of the anisotropy parameter through this method requires that the condition be fulfilled simple entertainment. As we will see, one way to do this is to estimate simultaneously the cross-section of absorption and extinction.


The paper is organized as follows. In Section II it is present the theoretical framework that supports the experimental measurements. In Section III shows the experimental arrangements. In Section IV presents the materials and methods used in this study. In Section V presents the results and discussion of measurements. Finally, in Section VI presents the conclusions.

\section{THE ANISOTROPY PARAMETER }\index{Factor de anisotrop\'ia}

The estimation of anisotropy ($ g $) normally involves measuring the angular distribution of the scattered light. However, we will see that this parameter can be estimated using integrating spheres.

For extremely diluted medium we can assume that the cells scatter from each other independently and there is no shading effects or multiple scattering. In this case, it is possible to match approximately the transmittance, reflectance and absorption properties of the medium with the cell anisotropy. Following is an analysis to justify this assertion.

Suppose we have a medium with particles having a phase function $ P (\theta, \phi) $. Must then be satisfied that
\begin{equation}\label{1}
 \int_{4\pi}P(\theta,\phi)d\Omega=1.
\end{equation}
If this particle distribution produces a pattern of isotropic scattering around $ \phi $, we have
\begin{equation}\label{2}
2\pi\int_{0}^{2\pi}P(\theta)\sin\theta d\theta=1.
\end{equation}
With the change of variable $ x = \cos \theta $ the integral is simplified and we can write
\begin{equation}\label{3}
 \int_{-1}^{1}p(x)dx=1,
\end{equation}
where the function $p(x)=2\pi P(\cos^{-1}x)$.

The anisotropy parameter is defined as the first moment of the phase function. That is,
\begin{equation}\label{4}
g=\int_{-1}^{1}xp(x)dx. 
\end{equation}
Since $ x = \cos \theta $, $ g $ represents the average cosine of the scattering angle (as already mentioned).

The fraction of scattered light that is scattered forward is given by,
\begin{equation}\label{5}
 f=\int_{0}^{1}p(x)dx.
\end{equation}
Similarly, the fraction of scattered light exiting backwards is
\begin{equation}\label{6}
b=\int_{-1}^{0}p(x)dx.
\end{equation}
and should to comply that $ f + b = $ 1.

We write now,
\begin{equation}\label{7}
 g=g_f+g_b
\end{equation}
where
\begin{equation}\label{8}
 g_f=\int_{0}^{1}xp(x)dx
\end{equation}
and
\begin{equation}\label{9}
 g_b=\int_{-1}^{0}xp(x)dx.
\end{equation}

We consider first the case of large particles. In this case, the spread is primarily at small angles, either forward or backward. For these angles $ x = \cos \theta \simeq \pm 1$ , so $ g_f \simeq f $ and $ g_b \simeq-b $.

With this approach we need
\begin{equation}\label{10}
 f-b=g.
\end{equation}
Using the condition,
\begin{equation}\label{11}
 f+b=1,
\end{equation}
and solving for $ f $ and $ g $, we have that the fraction of light scattered forward can be written as
\begin{equation}\label{12}
 f=\dfrac{1+g}{2},
\end{equation}
while the fraction of light scattered backwards is,
\begin{equation}\label{13}
 b=\dfrac{1-g}{2}.
\end{equation}

We can see that a value of $ g = 1$ (all light scattered forward) results in $ f = 1$ and $ b =0$. Furthermore, with $ g =  -1$ (all the light scattered backwards) we obtain $ f = 0 $, and $ b =  1$. This is consistent with expectations.

Furthermore, for small particles it has isotropic scattering. In this case, $ p (x) = p (-x) $, $ g = 0 $, and $ f = b = 1/2 $. We see that this result also agrees with the approximation (\ref{12}) and (\ref{13}). This suggests that those who were motivated by an approximation for large particles, may be reasonable approximations for particles of other sizes. This suggests that these approaches, who were motivated by an approximation for large particles, may be reasonable to particles of other sizes.

\begin{figure}[t]
    \centering
 \includegraphics[width = 7cm]{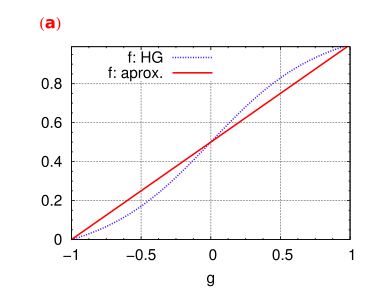}
\includegraphics[width = 6.75cm]{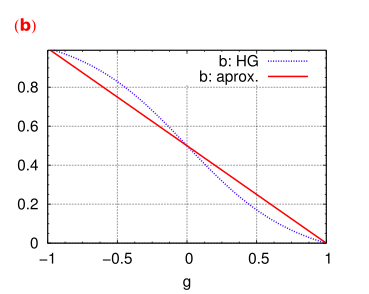}
 
 \vspace{.2cm}

   \caption{Comparison between the values of $ f $ and $ b $ with approximation $ x \simeq \pm $ 1 (straight line) and the phase function of Henyey-Greenstein.\label{fig:HG_fb}}
\vspace{1cm}
\end{figure}

It is clear that the validity of these expressions for intermediate values of $ g $ depends on the phase function, but it is interesting to explore if the phase function of Henyey-Greenstein. Figure \ref {fig:HG_fb} shows the behavior of the values of $ f $ and $ b $ calculated using the expressions (\ref {12}) and (\ref {13}) (solid line) and calculated based on the phase  function of Henyey-Greenstein (dotted line). As expected, the curves coincide in the limit $ g = -1 $ y $ g = 1 $, as well as the center point $ g =  0$. There are only small deviations for other values of $ g $. This gives some confidence in the approach represented by the expressions (\ref {12}) and (\ref {13}), which also implies that $ g = f-b$.

This shows the possibility of estimating the anisotropy factor $ g $ in terms of the properties of reflectance and transmittance of the diluted medium. That is, without requiring angle scattering measurements.

We know that, by definition, absorption ($ A $), reflectance ($ R $) and transmittance ($ T $) satisfy

\begin{equation}\label{14}
 {\cal{A}}+{\cal{R}}+{\cal{T}}=1.
\end{equation}
Another way to write this relationship is,
\begin{equation}\label{15}
 \dfrac{{\cal{T}}}{1-{\cal{A}}}+\dfrac{{\cal{R}}}{1-{\cal{A}}}=1.
\end{equation}
The reason for doing this is to make clear that this is equivalent to the condition $ f + b = 1 $, so we have to
\begin{equation}\label{16}
 f= \dfrac{{\cal{T}}}{1-{\cal{A}}}, \hspace{2cm} b= \dfrac{{\cal{R}}}{1-{\cal{A}}}.
\end{equation}

Hence, it is possible to determine the value of $ g $ through measurements of transmittance and reflectance coefficients. Thus from equations (\ref{10}) and (\ref{16}),
\begin{equation}\label{17}
g=\dfrac{{\cal{T}}-{\cal{R}}}{{\cal{T}}+{\cal{R}}}. 
\end{equation}

Note that for this we are assuming that media is sufficiently diluted as to have only single scattering, plus you can collect all the transmitted and reflected light.

We can say that since we have $ {\cal R} $, $ {\cal T} $ and $ {\cal A} $, with the same arrangement we can estimate the parameter of anisotropy of the cells by dilutions of the medium. Although in principle we would have spurious contributions in reflectance due to air-interface cell plastic we can correct it by using a reference cell and removing the specular component measurements.

The arrangement allows us to estimate the absorption  ($ \cal A $) from the reflectance measurements ($ \cal R $) and transmittance ($ \cal T $) of the medium, as mentioned above. 

Once we have estimated absorption, we used the relationship between this and the absorption cross section $C_a$ of the particles,
\begin{equation}\label{19}
 {\cal A}=\exp(-\rho C_a L),
\end{equation}
where $ {\cal A} $ is the ratio of the the  reference power $P_o$  (measured in the absence of particles) and the   reduced power  $P_m'$ (with the particles present),  $ \rho $ is the density of particles,  and $ L$ is the thickness of the cell [see equation (4) from Ref. \cite{teran:10}].  Solving for $ C_a$, we have,
\begin{equation}\label{20}
 C_a=-\dfrac{1}{\eta}\ln({\cal A}),
\end{equation}
where $ \eta = \rho L $ represents the number of particles per projected area.

 As we will see in the next section  the estimation of this parameter allows us to determine the anisotropy of the particles adequately.

\section{Experimental setup}

\begin{figure}[t]
\centering
\includegraphics[scale=.5]{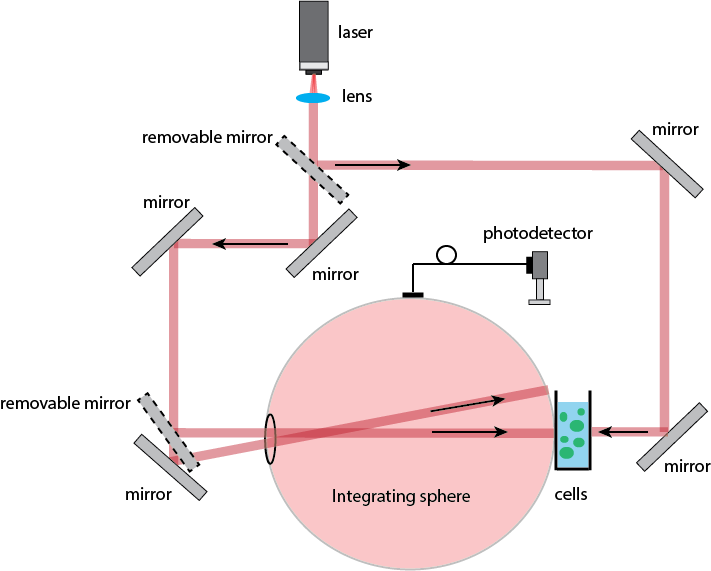}  
\hspace{.1cm}  
  
\caption{Scheme of the experimental setup to estimate the parameter of anisotropy of micro-particles. Figure (a) shows the configuration to determine the proportion of light scattered forward $ f $. Figure (b) shows the configuration to determine the proportion of light scattered backward $ b $.
\label{fig:esfint}
}
\end{figure}

The diagram in Figure \ref {fig:esfint} shows the way in which we estimate the amount of scattered light using an integrating sphere as a collector of light, a quartz cell with the particle suspension and a photodetector for measurement. It is noteworthy that for particles with diameters in the micron range most of the scattering is forward, so we can consider that the light returning from the field to the sample does not influence the measurements, which simplifies the estimation of the amount reflected or transmitted by the suspension of light particles.

The arrangement consists of an integrating sphere Ocean Optics (Model ISP-50-8-R-GT), a NewPort (Model 2051-FS-M) photodector, a laser HeNe ($ 12 $ mW, random polarization) a quartz cell with $ 1 $ cm thick, 5 permanent mirrors and $ 2 $ removable.  The right arm of the array illuminates the particles in transmittance and the left arm in  reflectance. Note that disabling the removable mirror 1, the beam illuminates the sample directly to an angle $ 8 ^ {\circ} $, the integrating sphere has an aperture placed in this position to pass up the specular component. On the other hand, the beam 3 also illuminates the sample at an angle $ 8 ^ {\circ} $ transmittance but in this case we also let the specular component leave  the sphere. The integrating sphere employed (Labsphere 4P-GPS-053-SL) has a wall reflection coefficient $m =0.99$, an internal diameter of $13.46$cm, two $2.54$cm diameter ports, and a fiber-coupled exit port in which we connected the fiber optics photodetector (?). 

The photodetector will collect a detected power $P_d$  when  illuminate with  the beam $ 1 $. Then  if we illuminate with the beam $2$ we will get a  power $ P_r$.  However, if we illuminate with the beam $3$ we will get a  power $ P_t$. From Ref. \cite{teran:10} [see equations (8) y (9)], we have that the reflectance and transmittance can be written as
\begin{equation}\label{24}
{\calÊR} = m\dfrac{P_r}{P_d}
\mbox{\hspace{1cm}and\hspace{1cm}} 
{\cal T} = m\dfrac{P_t}{P_d}.
\end{equation}

From the  equations (\ref {17}) and (\ref {24}),  we have that the anisotropy parameter can be written as,
\begin{equation}\label{25}
g = \dfrac{P_r-P_t}{P_r+P_t}.
\end{equation}


The method was applied to a sample of powdered titanium dioxide (TiO$ _2 $). The main advantage of working with this kind of particles is that we can characterize optically with Mie theory, which allow us to check the consistency of the experimental measurements. 

\begin{figure}[t]
\centering
\includegraphics[scale=.6]{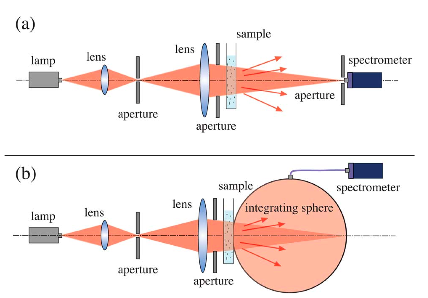}  
\hspace{.1cm}  
  
\caption{Schematic diagrams of the arrangements used for the characterization of the photosynthetic cells. Setup for the measurement of (a) extinction and (b) absorption \cite{teran:10}.
\label{fig:ae}
}
\end{figure}

Another way to check the consistency of the experimental measurements is through the estimation of the absorption and the extinction of the sample.  In order to do this we use the methodology presented in Ref. \cite{teran:10}. The  arrangement of the Fig. \ref{fig:ae}(a)  allows us to estimate the extinction cross section $C_e$ through the expression
\begin{equation}\label{26}
 { P_m}=\exp(-\rho C_e L),
\end{equation}
where $P_o$ is the reference power (without particles), $P_m$ the power collected by the photodetector (with a suspension of particles),  $ \rho $ is the density of particles,  and $ L$ is the thickness of the cell. On the other side, the Fig. \ref{fig:ae}(b)  allows us to estimate the absorbtion cross section $C_a$ through the expression
\begin{equation}\label{27}
 { P_m'}=P_o\exp(-\rho C_a L)
\end{equation}
where  $P_m'$ the power collected by the photodetector (with a suspension of particles). We should note that although the equations (\ref{26}) and (\ref{27}) are very similar estimation of the values of $P_m$ and $P_m'$ are very different. For a deeper description of the estimation of these parameters see Ref. \cite{teran:10}.

We did dilutions of the sample to verify the linear dependence between the number of particles per area projected $ \eta $ and the cross section of absorption $ C_a $, which is true only  if the condition of independent  and single scattering are valid.  Therefore, we consider that the measurement region for estimating valid absorption cross sections and extinction of particles is in the range where the measurements follow a linear behavior as a function of the dilutions, see Fig. \ref {fig:Cae_TiO2}.

\section{Results and Discussion}

In this section we estimate the anisotropy of the sample of TiO2 and we compare this value with that obtained with the Mie theory. The sample of TiO$_2 $ that we consider is composed of particles of an average diameter $ 0.25 \ \mu $m, an index of refraction of particle $n_T =  2.093 $ immersed in water. In this case the Mie theory leads us to obtain an anisotropy parameter $ g =  0.59$ and a extinction  cross section of $C_e = 0.068  \ \mu$m$^ 2$ to  $\lambda = 632.08 \  $nm.

\begin{figure}[t]
\centering
\includegraphics[scale=.5]{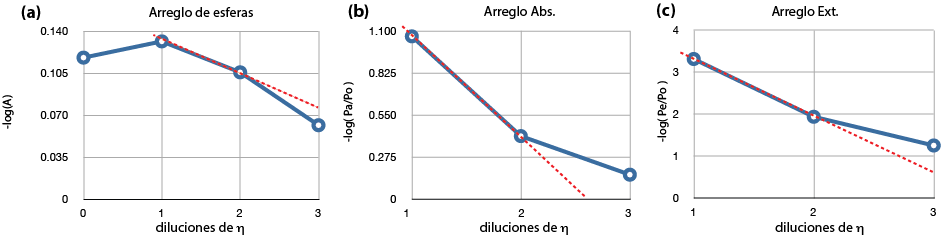}   
\caption{
Par\'ametro de anisotrop\'ia de polvo de TiO$_2$ de part\'iculas de $0.25\ \mu$m de di\'ametro para diferentes diluciones.
\label{fig:Cae_TiO2}
}
\end{figure}

We will start our measurements ensuring that the condition of single scattering is satisfied. So it was necessary to determine the absorption and extinction spectrum  of the sample.  Experimental measurements of absorption and extinction of TiO$_2$  are shown in Figs. \ref {fig:Cae_TiO2}.  The Fig.  \ref {fig:Cae_TiO2}(a) was obtained with the arrangement of the integrating sphere Fig. \ref{fig:esfint} and Figs. \ref{fig:Cae_TiO2} (b) and \ref{fig:Cae_TiO2} (c) the absorption and extinction arrangements shown in Figure \ref{fig:ext_abs}, respectively. In this case four sample dilutions were made, dilution $ 0 $ represents the original medium, dilution means $ 1 $ $ 2 $ diluted in parts and so on. We can see that to  dilution $ 0 $ the media was highly concentrated so that we started  from the dilution $ 1 $ --see Figs. \ref{fig:Cae_TiO2}(b) and \ref{fig:Cae_TiO2} (c)-- and ended up in the third dilution because the environment was very diluted. The dotted red line represents a least squares fit of the selected points.

The absorption cross section $C_a$ of  TiO$_2 $ can be estimated through a least squares fit to the points shown in Fig. \ref{fig:Cae_TiO2}(a) and equations(\ref{20}),
\begin{equation}\label{28}
C_a^{(1)}=1.16\times 10^{-3} \ \mu\mbox{m}^{-2}.
\end{equation}
On the other hand, from Fig. \ref{fig:Cae_TiO2}(b) we have that the absorption cross section obtained with absorption arrangement is Ref. \cite{teran:10},
\begin{equation}\label{29}
C_a^{(2)}=2.95\times 10^{-2} \ \mu\mbox{m}^{-2}.
\end{equation}
where the number of particles per projected area 
\begin{equation}\label{30}
\eta=96.6\ \mu\mbox{m},
\end{equation}
was obtained from the experimental measurements of extinction \mbox{[Fig. \ref{fig:Cae_TiO2}(c)]} and the extinction cross section $C_e$ obtained with Mie theory.

 We can see that there is a significant difference (one order of magnitude) in the estimation of this parameter, if we use the integrating sphere arrangement or arrangement of absorption. This is due to that the arrangement of the sphere captures the light coming forward and back, while the arrangement of absorption  captures light only going forward, which leads us to have an overestimation of $C_a$, as we can see from the values shown in equations (\ref{27}) and (\ref{28}). However, for more absorbent or larger particles  the difference is not so marked, because there will be less light scattered backwards, especially for spherical particles, however, for particles with other geometries the situation may change and you need to take into account the limits of the array of absorption.
 
\begin{figure}[t]
\centering
\includegraphics[scale=.7]{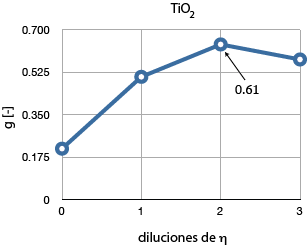}   
\caption{
Anisotropy of TiO $ _2 $  as a function of the dilution of the media. 
\label{fig:g_TiO2}
}
\end{figure}

The anisotropy of the particles of TiO$_2 $ was obtained from the values shown in Figure \ref {fig:g_TiO2}. The vertical axis shows the measurements obtained with the arrangement of the sphere (see Figure \ref{fig:esfint}) and equation (\ref{eq:gg}), the horizontal axis the dilutions of the media. The  dilution $0$ corresponds to the media undiluted where $ \eta = 99.6 \ \mu\mbox {m} ^{-2}$, each dilution represents the third of the above, so that dilution for 2 we have $ \eta/ 9 = 11.06 \ \mu$m$^{-2}$.

We can see that the variations in the anisotropy parameter is related to the cases where we have many or too few particles. Remember that the anisotropy parameter is valid only when we have single sacttering, like $C_a $, so that the region where the latter is valid it will also $ g $. Thus, in Fig \ref{fig:g_TiO2}, we will consider as valid those points where the value of $ g $ is constant or take their greatest value this case corresponds to dilution $2$, so that the anisotropy of TiO$_2$  is,
 \begin{equation}\label{31}
 g=0.615.
 \end{equation}

\begin{table}[b]
\caption{
Propiedades \'opticas del TiO$_2$.
\label{tab:musres}
}
\begin{center}
\begin{tabular}{|c|c|c|c|}

\hline
&\mbox{\hspace{.5cm}}$C_e$ [$\mu$m$^2$]\mbox{\hspace{.5cm}}&$\hspace{.5cm}C_a$ [$\mu$m$^2$]\mbox{\hspace{.5cm}}&\hspace{1cm}$g$ [-] \mbox{\hspace{1cm}}\\\hline
\hline
Experimental&-&$1.16\times 10^{-3}$ &0.615\\\hline
Calculado&0.068992&-&0.59945\\\hline

\end{tabular}
\end{center}
\label{default}
\end{table}%

As we have seen it is very important to ensure that the condition of  single scattering  is fulfilled. One way to know this is through the product of $ \eta \times  C_s$, which gives us an estimate of the "area" occupied by the particles in the illuminated region. For values of $ \eta  C_s$ near $ 1 $ or higher imply that multiple scattering events occur, while for very small values $ \eta C_s \ll  1$ will have single scattering. In the case, we have $ \eta / 9 \times C_s = (99.6 / 9) \times (0.06684) =  0.73$, which tells us that dilution $2$ is a good approximation to  single scattering.

\section{Final remarks}

The results show that for the particles of $TiO _2 $ the experimental and theoretical results are consistent. However, the borosilicate glass powder has inconsistencies between these results. This may be because in principle we would have spurious contributions in reflectance due to the air-plastic cells, we could be corrected using a reference cell.


\begin{thebibliography}{xx}
\bibitem{Barthelemy:08} Barthelemy, Pierre and Bertolotti, Jacopo and Wiersma, Diederik S., "A Levy flight for light", Nature, 2008,Vol., 453, Number 7194, 495--49
\bibitem{chan:43} S. Chandrasekhar, "Stochastic Problems in Physics and Astronomy", Rev. Mod. Phys. 15, 1Ð89 (1943)
\bibitem {pereira:04} Pereira,E.,Martinho,J.M.G., and Berberan-Santos, M.N. 2004. ''Photon trajectories in incoherent atomic radiation trapping as L\'evy flights". Physical Review Letters, 93,
120201
\bibitem {odonell:82} K. O'Donnell, "Speckle statistics of doubly scattered light," J. Opt. Soc. Am.  72, 1459-1463 (1982).
\bibitem{teran:10} E. Ter\'an, E. M\'endez, S. Enr\'iquez, and R. Iglesias-Prieto, "Multiple light scattering and absorption in reef-building corals," Appl. Opt.  49, 5032-5042 (2010).

\bibitem{wang:95} Wang, L-H, S.L. Jacques, L-Q Zheng: MCML - Monte Carlo modeling of photon transport in multi-layered tissues. Computer Methods and Programs in Biomedicine 47:131-146, 1995.

\bibitem{1} Determination of g and mu Using Multiply Scattered Light in Turbid Media  
\bibitem{2} Backscattered stokes vectors of turbid media: anisotropy factor and reduced scattering coefficient estimation

\end{thebibliography}
\end{document}